\documentclass[12pt,preprint]{aastex}

\def\arcsecpoint{$''\!.$}

\shorttitle{Obscuration Towards the NLR in Seyfert Galaxies}
\shortauthors{Kraemer et al.}

\begin{document}

\title{Multi-wavelength Probes of Obscuration Towards the Narrow Line Region
in Seyfert Galaxies} 

\author{S. B. Kraemer\altaffilmark{1}, H.R. Schmitt\altaffilmark{2},
D. M. Crenshaw\altaffilmark{3}, M. Mel\'endez\altaffilmark{4}, 
T.J. Turner\altaffilmark{5}, M. Guainazzi\altaffilmark{6}, \& R.F. 
Mushotzky\altaffilmark{7}}

\altaffiltext{1}{Institute for Astrophysics and Computational Sciences,
Department of Physics, The Catholic University of America, Washington, DC 20064; and
Code 667, Astrophysics Science Division, NASA Goddard Space Flight Center, Greenbelt,
MD 20771.}
\altaffiltext{2}{Remote Sensing Division, Naval Research Laboratory,
Washington, DC 20375; and Computational Physics Inc., Springfield, VA 22151.}
\altaffiltext{3}{Department of Physics and Astronomy, Georgia State University,
Astronomy Offices, Atlanta, GA 30303.}
\altaffiltext{4}{Henry A. Rowland Dept. of Physics \& Astronomy, The Johns Hopkins University, Homewood Campus, Baltimore, MD
21218} 
\altaffiltext{5}{Dept. of Physics, University of Maryland Baltimore County,
Baltimore, MD 21250}
\altaffiltext{6}{European Space Astronomy Centre of the European Space
Agency, PO Box 78, Villanueva de la Ca\~{n}ada, e-28691 Madrid, Spain} 
\altaffiltext{7}{Department of Astronomy, University of Maryland, College Park,
MD 20742}

\begin{abstract}

We present a study of reddening and absorption towards the Narrow Line Regions (NLR) in
active galactic nuclei (AGN) selected from the Revised Shapley-Ames, 12$\mu$m,
and {\it Swift}/Burst Alert Telescope samples. For the sources
in host galaxies with inclinations of $b/a > 0.5$, we find that mean ratio
of [O~III] $\lambda$5007, from ground-based observations, and
[O~IV] 28.59$\mu$m, from {\it Spitzer}/Infrared Spectrograph 
observations, is a factor of 2 lower in Seyfert 2s than Seyfert 1s. 
The combination of low [O~III]/[O~IV] and [O~III] $\lambda$4363/$\lambda$5007 ratios
in Seyfert 2s suggests more extinction of emission from the NLR than in Seyfert 1s. Similar column
densities of dusty gas, N$_{H} \sim $ several $\times$ 10$^{21}$ cm$^{-2}$, can account for 
the suppression of both [O~III] $\lambda$5007 
and [O~III] $\lambda$4363, as compared to those observed in Seyfert 1s.
Also, we find that the X-ray line O~VII $\lambda$22.1 \AA~ is
weaker in Seyfert 2s, consistent with absorption by the same gas
that reddens the optical emission. Using
a {\it Hubble Space Telescope}/Space Telescope
Imaging Spectrograph slitless spectrum of the Seyfert 1 galaxy NGC 4151, we estimate
that only $\sim$ 30\% of the [O~III] $\lambda$5007 comes from within 30 pc of
the central source, which is insufficient to account for the low [O~III]/[O~IV] ratios
in Seyfert 2s. If Seyfert 2 galaxies have similar intrinsic [O~III] spatial profiles, the external dusty gas 
must extend
further out along the NLR, perhaps in the form of nuclear dust spirals
that have been associated with fueling flows towards the AGN.

\end{abstract}

\section{Introduction}

Seyfert galaxies are relatively moderate luminosity ($L_{bol} \lesssim 10^{45}$ ergs$^{-1}$), local ($z<0.1)$) AGN.
Historically, they have been classified based on their optical spectra, with Seyfert 1s
possessing broad permitted lines (full width at half maximum [FWHM] $\gtrsim$ few 1000 km s$^{-1}$), narrower forbidden
lines (FWHM $\lesssim$ 1000 km s$^{-1}$) and non-stellar continua, and Seyfert 2s possessing
permitted and forbidden lines of similar widths and optical continua dominated by the host
galaxy (Khachikian \& Weedman 1974). The discovery of polarized broad permitted lines and non-stellar
continua in spectro-polarimetric observations of Seyfert 2 galaxies (e.g., Miller \& Antonucci 1983) led to the unified model
for Seyfert galaxies (Antonucci 1993), which posits that the central AGN is surrounded by a dusty,
molecular torus and the difference between the two types is due to our line of sight
with respect to the torus. Specifically, our view of the broad emission line region and central continuum
source in Seyfert 2s is blocked by the torus in the above picture. 

The forbidden lines detected in Seyfert galaxies arise from the so-called narrow-line region (NLR) which may 
extend from 1pc to several hundred pcs from the central source. Based on
narrow-band [O~III] images, originally from ground-based observations (Pogge 1988a,b; 1989) and, subsequently images obtained
with the {\it Hubble Space Telescope (HST)} (Schmitt \& Kinney (1996), the NLR in many Seyferts
exhibits a bi-conical structure, consistent with collimation of the ionizing
radiation emitted by the central source, either by the torus or an optically thick wind emanating from the
accretion disk (e.g. K\"{o}nigl \& Kartje 1994). Depending on the scale height of the torus, one
would expect that the Seyfert 1s and 2s would exhibit similar NLR emission-line properties, while their
structure should depend primarily on viewing angle, with Seyfert 1s being more symmetric and compact. While the
latter is generally the case (Schmitt et al. 2003a), with some caveats (e.g. Mulchaey et al. 1994), there are clear differences
in the emission-line spectra. Nagao et al. (2001) have shown that the density-sensitive [O~III] $\lambda$4363/$\lambda$5007
ratio (hereafter, R$_{\rm O3}$) is greater in Seyfert 1s, indicating the contribution from gas with density $n_{H}$
$>$ 10$^{5.5}$ cm$^{-3}$, the critical density for de-excitation of the $^{1}D_{2}$ level of O~III (Osterbrock \& Ferland 2006). 
Furthermore, high ionization lines such as [Fe~X] $\lambda$6374
are relatively stronger in Seyfert 1s. Nagao et al. speculate that the dense [O~III]-emitting region and the high ionization
gas are in the inner part of the NLR, and hence may be obscured by the torus in Seyfert 2s. The relationship
between the FWHM of emission lines, detected in the spectra of Seyfert galaxies, and the critical densities
of their upper levels (e.g.,
Filippenko \& Halpern 1984; de Robertis \& Osterbrock 1986), is further evidence for density stratification in the NLR. 

The putative torus, however, may not be the only source of extinction towards the emission-line regions
in Seyfert galaxies. There appears to be dust in our line-of-sight towards the NLR of Seyfert galaxies, with
reddening of $E_{B-V} \sim 0.2 - 0.4$ mag (e.g., Cohen 1983;  MacAlpine 1988; Ferland \& Osterbrock 1986; Kraemer et al. 1994).
In our analysis (Kraemer \& Crenshaw 2000a) of long slit spectra of the NLR of the Seyfert 2 galaxy NGC 1068, obtained with
{\it HST}/Space Telescope Imaging Spectrograph (STIS), we found extinction of the optical and UV emission lines consistent with a ``screen'' of dust\footnote{The concept
of a screen is an idealization, with the more likely scenario being that the dust is in the plane of the host galaxy
or in the form of dusty nuclear spirals. Note, all quoted column densities in this paper are for the simple case of
uniform screen, external to the NLR.}, external
to the optical NLR, with a hydrogen column density N$_{H}$ $\sim$ 10$^{21}$ cm$^{-2}$, assuming a Galactic dust-to-gas
ratio (Shull \& van Steenberg 1985). On the other hand, while there is less evidence for an external screen
in the Seyfert 1 galaxy NGC 4151 (Kraemer et al. 2000), near IR emission detected in Gemini/Near-Infrared Integrated Field Spectrograph
(NIFS) observations
revealed that there is considerable material outside the optical emission-line bicone (Storchi-Bergmann et al. 2009).
Furthermore, photoionization models of the NLR emission-line gas in both of these objects, as well
as the Seyfert 2 galaxies Mrk 3 (Collins et al. 2009) and Mrk 573 (Kraemer et al. 2009) require a contribution from
low-ionization gas ionized by a heavily absorbed continuum, hence, outside the NLR mapped in the light of [O~III]. If
this low-ionization gas is dusty and along our line-of-sight to the NLR, emission lines from within the
optical NLR, such as [O~III], would be reddened.  
There is also evidence for dust features in the inner nuclei of Seyfert galaxies, possibly associated with 
fueling flows (Martini \& Pogge 1999; Martini et al. 2003a; Crenshaw et al. 2010a), which would be a strong source
of extinction if they crossed our line-of-sight to the NLR. 

Although optical emission-lines, such as [O~III] $\lambda$5007, may be reddened by gas
outside the NLR, mid-IR emission lines such as [O~IV] 25.89 $\mu$m will be much less affected. In comparing
the ratio of [O~III], from ground-based observations, [O~IV] from {\it Spitzer}/IRS spectra, among the 9-month sample 
of AGN detected by the {\it Swift}/Burst Alert Telescope
(BAT) (Tueller et al. 2008), Mel\'endez et al. (2008a) found that Seyfert 2s had relatively weaker [O~III] than Seyfert 1s\footnote{Although
weak [O~III] could be an indication that ionizing photons are absorbed before they reach the NLR, as suggested by Trouille \& Barger (2010),
in such a scenario the [O~IV] should be similarly affected, which is not typically the case (although see discussion in
Weaver et al. [2010]).}, consistent
with an N$_{H}$ $\approx$ 1--10 $\times$ 10$^{21}$ cm$^{-2}$. This effect was confirmed
by Diamond-Stanic, Rieke, \& Rigby (2009) and by Baum et al. (2010), using Seyferts in the revised Shapley-Ames, comprised of sources from Maiolino
\& Rieke (1995) and Ho, Filippenko, \& Sargent (1997), and
12$\mu$m (Rush, Malkan, \& Spinoglio 1993) samples, respectively. Diamond-Stanic et al. note that the effect is present even after correcting
the [O~III] fluxes using the Balmer decrement, which suggests that there are regions from which the
optical emission is essentially undetectable. Baum et al. proposed that the unobservable gas lies behind the molecular torus
(see, also Nagao et al. 2001). Dudik et al. (2007) found that the ratio of the
[Ne~V] 14.3 $\mu$m/24.3 $\mu$m lines in
a number of AGN was below the low density limit, hence consistent with extinction. They also suggest that some fraction
of the [Ne~V] emission arises from behind the torus.  However, based on photoionization
models, Mel\'endez et al. (2008a) determined that the [O~IV] emission region was $\sim$ a few 10s of pcs from the
AGN for most of the targets in the 9-month BAT sample, which requires that the tori possess very large scale heights.
Nevertheless, taking the evidence from the R$_{\rm O3}$ and the [O~III]/[O~IV] ratios together, one might conclude that
the extinction is heaviest towards the inner regions of the NLR. 

One open question is whether the extinction is due to dust associated with the AGN, either the torus or the NLR, or
dust in the host galaxy. Keel (1980) found a deficiency of Seyfert 1s in nearly edge-on hosts and suggested that 
Seyfert galaxies have larger bulges or thicker disks than normal spiral galaxies. Maiolino \& Rieke (1995) found more Seyfert 1.8s and 1.9s
in edge-on systems, hence the reddening of emission from the BLR, which would suppress the broad Balmer lines, could be due to dust in the 
plane of the host galaxy. Schmitt et al. (2001) also noted a lack of Seyfert 1s in edge-on systems. 
Hence, one
aim of this paper to to isolate
the effect of reddening due to gas associated with the AGN, rather than within the plane of the host galaxy.
Beyond that, we will determine how much of the [O~III] emission is undetected in Seyfert 2s, and, finally, constrain the column densities
of the gas outside the optical emission line bicone.
 
\section{Sample Selection and the Effect of Inclination}

In order to determine if there are significant differences in the [O~III]/[O~IV] ratios of Seyfert galaxies, we
require a sample that includes a sufficient number of Seyfert 1s and 2s that 
our results are not biased, e.g., by small number statistics.
To begin with, we included AGN for the revised Shapley-Ames sample
with IRS spectra analyzed by Diamond-Stanic et al. (2009) and from the 12$\mu$m sample, with IRS spectra analyzed by
Tomassin et al. (2008, 2010). Since both of these samples are weighted towards Seyfert 2s, we added
sources from the 22-month BAT sample (Tueller et al. 2010) with IRS spectra analyzed by
Mel\'endez et al. (in preparation). We stress the point that this is
not intended to be a complete, unbiased sample. However, as we will show,
the range in [O~III]/[O~IV] ratios and the dependence of this ratio on Seyfert
classification is similar to that observed in other studies. 

One goal of this study is to isolate the effects of dust associated with the host galaxy from that
associated with the NLR. The first step was to eliminate from our sample merging systems and those designated as 
residing in peculiar host galaxies, based on the galaxy morphologies from the RC3.9
(de Vaucouleurs et al. 1991). The remaining AGN include: 53 Seyfert 2s,
30 Seyfert 1s, 15 Seyfert 1.8/1.9s (``intermediate'' Seyferts), and 8 LINERs (see Tables 1 -- 4).
The uncertainties in the [O~IV] fluxes are found in the references cited in the tables, but are
generally $\sim$ 10\%.
While uncertainties in the [O~III] fluxes are rarely given,
Peterson et al. (1991) find that seeing variations in typical ground-based 
spectroscopy of AGN can result in photometric errors as large as 20\%. We 
therefore assume that our [O~III] fluxes are uncertain at this level. In Figure 1, we
show the [O~III]/[O~IV] ratio as a function of the [O~IV] luminosity, which can
be used as a proxy for the intrinsic luminosity of the AGN (Mel\'endez et al.
2008a; Mel\'endez, Kraemer, \& Schmitt 2010). As the plot shows, the sample smoothly spans more than four orders of
magnitude in luminosity, hence we do not appear to be introducing obvious
biases by including sources from different parent samples (in particular, those
from the BAT sample, which includes more distant and luminous sources, see
Mel\'endez et al. [2008a]). It does appear,
however, that lower luminosity objects, primarily LINERS, have somewhat
higher [O~III]/[O~IV] ratios. As discussed in Netzer (2009), since 
the AGN in LINERs are weak, they cannot produce as much O$^{+3}$ as Seyferts.
Furthermore, starlight may be the dominant
form of ionization in LINERs (Cid Fernandes et al. 2009), which would
also result in higher [O~III]/[O~IV] ratios. Indeed, the 
one Seyfert 1 in the low-ionization, high
[O~III]/[O~IV] region, Mrk 352, has [Ne~II] 12.81$\mu$m/[Ne~III] 15.56
$\mu$m $\sim$ 1.6 (Weaver et al. 2010), indicative of strong star
formation (e.g., Mel\'endez et al. 2008b), which could enhance [O~III].

From their study of {\it Spitzer}/IRS spectra of Seyfert galaxies,
Deo et al. (2007) found that the few objects in their sample that showed
deep 10 $\mu$m silicate absorption were in highly inclined or merging
systems. Crenshaw \& Kraemer (2001) determined that the UV continuum
reddening in Seyfert 1s increases dramatically with the inclination of the host
galaxy. This suggests that much of the extinction is due to dust the plane
of the host galaxy. 
In Figure 2
we show the effect of inclination in the sample by plotting the [O~III]/[O~IV] ratios
versus the axial ratios ($b/a$) of their host galaxies. 
Following Crenshaw \& Kraemer (2001), we have taken $b/a$ from (in order of
preference) de Zotti \& Gaskell (1985), Kirhakos \& Steiner (1990), and
the NASA/IPAC Extragalactic Database (NED) (see Tables 1--4). While there is no
strong correlation between [O~III]/[O~IV] and $b/a$ for the sample, 
there are clearly relatively fewer sources with [O~III]/[O~IV] greater than unity 
in inclined host galaxies. Therefore, in examining the effect of extinction
associated with the NLR, we opt to only consider those Seyfert galaxies in our sample
in hosts with $b/a > 0.5$ (see, also, discussion in Crenshaw \& Kraemer 2001).
Our final sample includes 40 Seyfert 2s and 26 Seyfert 1s. In Figure 2, one
can clearly see that Seyfert 1s, on average, have higher [O~III]/[O~IV] ratios
than do Seyfert 2s. Due to the ionization
effect discussed above, we do not include LINERs in the analysis. Intermediate Seyferts
exhibit a range of properties (e.g. Goodrich 1995; Trippe et al. 2010), including cases of weak AGN/strong starformation, that can
affect [O~III] independent from extinction, hence we also exclude these
from the analysis of the [O~III]/[O~IV] ratios. However, they are included
in our comparison with R$_{\rm O3}$ and soft X-ray emission (see Section 4). 

One possible complication in comparing [O~IV] fluxes from {\it Spitzer} with [O~III] fluxes
from ground-based observations is the aperture effect. For example,
some Seyfert galaxies show [O~III] emission from an extended narrow line region (ENLR) up to several
kpc from the central source, (e.g. Pogge 1989). For an object at a distance of 10 Mpc,
the approximate minimum distance to the Seyfert 1 and 2s in the sample,
1 kpc corresponds to $\sim$ 20$^{''}$, which could be outside the IRS
long-low or long-high apertures, with shorter dimensions of 10\arcsecpoint5 and 11\arcsecpoint1, respectively (Houck et al. 2004), 
depending on the orientation of the slit. However, as noted above, based on photoionization models,
Mel\'endez et al. (2008a) determined that the bulk of the [O~IV] emission arises close to the AGN, hence it is
unlikely that the IRS spectra have missed much of the flux. On the other hand, the [O~III] fluxes
used here come from a number of different ground-based observers (see Tables 1 --4), who
employed slits of various dimensions. Nevertheless, as demonstrated by Schmitt et al. (2003b), the [O~III]
profiles obtained {\it HST}/Wide Field and Planetary Camera 2 narrow-band images of Seyfert galaxies were strongly
centrally peaked and, in most cases, there was a good agreement
between the {\it HST} [O~III] fluxes 
and values from ground-based observations (see their Figure 2).
However, it is instructive to examine cases of nearby Seyferts with bright ENLRs. For example,
in Figure 3, we show the radial
profile of [O~III] $\lambda$5007, derived from a {\it HST}/STIS slitless G430M spectrum of NGC 4151.
In this case, the bulk of the [O~III] lies within $\sim$ 2$^{''}$ of the central peak, hence 
it is reasonably certain that the ground-based observation by Ho et al. (1997), using a
2$^{''}$X4$^{''}$ aperture, captured most of the [O~III] flux.  
For NGC 1068, the [O~III] flux measured by Schmitt et al. (2003b), within 4$^{''}$ of the AGN, is
$\sim$ 60\% of the value obtained by Bonatto \& Pastoriza (1997) through a
12\arcsecpoint5X3\arcsecpoint2 aperture, again indicating that the emission is centrally peaked.
Given that these two NLRs are close 
and extended, it is unlikely that aperture effects play a major role for the 
other AGN in our sample. As another check, if there is, in general, 
significant [O~III] from the ENLR, one may
expect a trend towards higher [O~III]/[O~IV] ratios with distance. However,
as seen in Figure 4, the [O~III]/[O~IV] ratio is not correlated with 
distance, which confirms that there is no overall aperture effect within the sample. 

\section{Comparison of the [O~III]/[O~IV] Ratios}

As mentioned above, several studies have noted that Seyfert 2s have, on average,
lower [O~III]/[O~IV] ratios than Seyfert 1s, which is attributed to
greater extinction towards the NLR in the former. With our reduced sample,
we have attempted to account for the effects of  
peculiar morphologies or host galaxy inclination. In Figure 5, we show
the distribution of the ratio for the Seyfert 1s and 2s in the reduced sample, and the
separation of the two classes is quite apparent.

The number of objects in our reduced sample is marginally sufficient
for a Kolomogorov-Smirnov (K-S) test to return meaningful results.  The K-S
test yields a  $D$-statistic $=$ 0.55 and a probability of the null
hypothesis $p$ $=$ 8.9 $\times$ 10$^{-5}$, indicating that the two populations
are different. A two-dimensional Kuiper
test of our reduced sample returns $D$ $=$ 0.525 and $p = 2.8 \times 10^{-3}$,
confirming the result of the K-S test.

The mean [O~III]/[O~IV] for the Seyfert 1s (excluding Mrk 352, see above) is
2.01, with a 95\% confidence interval for the mean of 1.43 -- 2.58. For the
Seyfert 2s, the mean is 0.97 with a 95\% confidence interval of 0.60 -- 1.26.
The ratio of the means is roughly the same found by LaMassa et al.
(2010) in comparing Seyfert 2s from SDSS to Seyfert 1s in the revised
Shapley-Ames sample (Diamond-Stanic et al. 2009). In order
to determine the average excess extinction in Seyfert 2s, we compared
the [O~III]/[O~IV] ratios for each object to the mean value for
the Seyfert 1s. We assumed 
$R_{v}= 3.1$, the extinction law of Cardelli, Clayton, \& Mathis (1989), and a dust-to-gas ratio resulting in $E(B-V) =1$ for 
N$_{H} = 5.2 \times 10^{21}$ cm$^{-2}$ (Shull \& van Steenberg 
1985). There were four Seyfert 2s with ratios greater than the mean 
value for the Seyfert 1s, namely, MCG$-$03$-$58$-$7, Tol 1238$-$364, NGC 1358 and NGC 4507,
so we did not include these sources in estimating the extinction; we list
the estimates for N$_{H}$ for the rest in Table 5.
For the Seyfert 2s, we find an average reddening of 0.41$\pm$ 0.27 mag, with the 95\% confidence 
interval of 0.32 -- 0.50.  The mean value for N$_{H}$ $=$  
2.1 $\times 10^{21}$ cm$^{-2}$, with the 95\% confidence interval of  
1.6 -- 2.6 $\times 10^{21}$ cm$^{-2}$. The range is somewhat smaller
than that found by Mel\'endez et al. (2008a), which is likely due
to the rejection of Seyferts in inclined hosts.

One caveat regarding estimates of reddening based on the [O~III]/[O~IV] ratios
is that it is likely, when
the host galaxy is viewed at low inclination, that half of the NLR emission lies
behind the galactic disk (e.g. Crenshaw et al. 2010b; Fischer et al. 2010). 
Therefore, depending on the amount of dust in the disk,
there can be significant extinction of the [O~III]. Interestingly, Mrk 3 resides
in an S0 host galaxy (de Vaucouleurs et al., 1991), hence its relatively large [O~III]/[O~IV]
ratio of 1.62 may be in part due lack of dust in the 
galactic disk. 

Intermediate Seyferts exhibit weak Balmer lines either due to the inherent
weakness of the BLR emission, reddening of the BLR, or a combination 
of these effects (e.g. Trippe et al. 2010). If the extinction also affects
the NLR, they might be expected to have [O~III]/[O~IV] ratios
similar to Seyfert 2s. In our sample, there is one clear outlier, NGC 3982,
with an [O~III]/[O~IV] $=$ 10.0, which is likely affected by a strong 
contribution from star formation (Mel\'endez et al. 2008b). Dropping
NGC 3982, the mean [O~III]/[O~IV] for the intermediate Seyferts is
1.24, with a 95\% confidence interval of 0.38 -- 2.10. A K-S test
comparing this ratio for intermediate Seyferts with that for Seyfert 1s
yields $D = 0.36$ and $p = 0.25$, while comparing them with Seyfert 2s yields
$D = 0.20$ and $p = 0.87$. Hence, based on the [O~III]/[O~IV] ratios, intermediate Seyferts
straddle the range of extinction among the Seyfert 1s and 2s. While there are too few LINERs in
our sample to perform a K-S test, their mean [O~III]/[O~IV] is 4.79, which is
consistent with their weak AGN and lower-ionization NLRs .   
 
\section{Comparison with other Emission Lines}

\subsection{The R$_{\rm O3}$ and Obscuration of the Inner NLR}

As mentioned in Section 1, Nagao et al. (2001) suggested that the lower values of R$_{\rm
O3}$
in Seyfert 2s were due to the obscuration of high density/high emissivity
gas in the inner NLR, which they characterized with 
$n_{H} = 10^{7}$ cm$^{-3}$. They further suggested that the differences in R$_{\rm
O3}$'s
between Seyfert 1s and 2s could be explained by a 5--20\% contribution from this
component in the former compared to a 0--5\% contribution in the latter.
Similarly, in order to match the R$_{\rm O3}$ observed
in the bright, central emission-line knot in NGC 4151, we (Crenshaw \& Kraemer 2005) generated a multi-component
photoionization model, with the [O~III] $\lambda$4363 emission dominated by 
gas with $n_{H} = 10^{6.2}$ cm$^{3}$, at a distance of $\sim$ 2 pc from the central source.

In Figure 6, we plot R$_{\rm O3}$ against [O~III]/[O~IV] (given the small
number of sources for which R$_{\rm O3}$'s have been published, we included
Seyferts in inclined hosts). There is no apparent correlation, 
with a Spearman rank coefficient of 0.14 and
significance of deviation from zero of 0.39. However, by over-plotting R$_{\rm O3}$ for 20,000K
in the low density limit (Osterbrock \& Ferland 2006) and the mean
[O~III]/[O~IV] for the Seyfert 1s in our sample, one can see there are 4
distinct quadrants. Most of the Seyfert 1s lie in the upper two quadrants, 
indicating a contribution from dense gas to the R$_{\rm O3}$. There are 9  
Seyfert 1s in the
upper left, with [O~III]/[O~IV] below the the minimum value of the 95\% confidence interval.
For several of these, the weak [O~III] could be due to
reddening by an external dust that covers the NLR, as is likely the case for NGC 4235, which is
inclined, and, possibly, for NGC 6814, MCG$-$6$-$30$-$15, and ESO 140$-$G043 
which have reddened continua (Morris \& Ward 1988). On the other hand,  
most (12/19) of the Seyfert 2s
lie in the lower left quadrant. We interpret this
as an indication that the extinction is greatest towards the inner NLR,
which includes the dense [O~III]-emitting gas (but, see discussion in Section 5.). Note that there
are a number of Seyfert 2s that have [O~III]/[O~IV] similar to the
(presumably) reddened Seyfert 1s, yet have lower R$_{\rm O3}$, which
is consistent with dust surrounding the inner NLR in the former versus
a more uniform external screen in the latter. There is some overlap
between the two classes near the middle of the diagram, which could be a result
of a relatively weak contribution from the dense component in some Seyfert 1s 
and/or less extinction towards the inner NLR in some Seyfert 2s, which, in turn,
may be the an indication of patchy dust that allows
a more direct view of the dense component.

There are a few odd points in Figure 6. Two Seyfert 2s lie in the
upper right quadrant: NGC 424 and NGC 1358, The former has also been
classified as an intermediate Seyfert (Murayama, Taniguchi, \& Iwasawa 1998), hence 
it is not necessarily surprising that it has properties in common with
Seyfert 1s. Although Nagao et al.(2001) list an R$_{\rm O3}$ for NGC 1358 based
on emission line ratios from Phillips, Charles, \& Baldwin (1983), neither
Ho, Filippenko, \& Sargent (1995) nor Vaceli et al. (1997) reported a detection of
[O~III] $\lambda$4363, therefore, R$_{\rm O3}$ is uncertain. The one intermediate
Seyfert, NGC 7314, is found in the lower left quadrant. However, it is
has a $b/a = 0.43$, hence there is likely to be
extinction due to dust in the plane. Among the Seyfert 1s, the very low
R$_{\rm O3}$ for Mrk 6 may be somewhat suspect, as Nagao et al. averaged
values from  Koski (1978) and Cohen (1983) and the latter measured relatively
stronger 4363 emission. The ``prototypical'' Seyfert 1 galaxy, NGC 4151, occupies
the most extreme position in the lower right quadrant.  There is
evidence, however, that the extended NLR in NGC 4151 was ionized when the
central source was significantly more luminous than at present
(e.g.. Schulz \& Komossa 1993; Wang et al. 2010). Hence, there may
be a relatively large contribution from low density gas which
would have the effect of reducing R$_{\rm O3}$.

\subsection{Soft X-ray Emission}

High resolution X-ray spectra Seyferts, obtained with 
{\it Chandra} (e.g., Sako et al. 2000; Ogle et al. 2000) and
{\it XMM-Newton} (e.g. Kinkhabwala et al. 2002), have
revealed the presence of numerous soft X-ray (energies $<$ 2 keV) emission lines.
Based on {\it Chandra}/ACIS imaging, the X-ray emission line region appears to be roughly
coincident with the [O~III] emission (Sako et al.2000; Ogle et al. 2000; Zeng 2009),
and it is probable that both the X-ray and optical emission line gas are photoionized by
the central source (Bianchi, Guainazzi, \& Chiaberge 2006).
The strongest emission line in
the soft X-ray spectra of Seyfert galaxies is typically the forbidden
line (1s$^{2}$$^{1}$S$_{0}$ -- 1s2s$^{3}$S$_{1}$) of He-like oxygen,
O~VII-f 22.1\AA~.   
If the NLR is surrounded by dusty gas, which reddens the [O~III], one would expect that
the same gas would absorb the O~VII.

We have X-ray fluxes for 15 sources, obtained from {\it XMM-Newton}/reflection grating spectrometer 
(RGS) data (see Table 6). Those
fluxes listed as from ``this paper'' are from the set of archival RGS spectra
analyzed by Guainazzi \& Bianchi (2007); the details of the spectral
analysis are described therein. 
Due to the small number of sources for which OVII-f line fluxes available, 
we have added NGC 424, NGC 7582, NGC 3227, which are in inclined
hosts, and NGC 2992, which was
left out of the [O~III]/[O~IV] comparison since it is in a peculiar  
host galaxy.
In Figure 7, we show the flux ratios O~VII-f/[O~IV] versus those of [O~III]/[O~IV].  
There is a clear trend, although with the caveat
that the sample size is small, with a Spearman rank coefficient of 0.736 and
significance of deviation from zero of 0.0008.
The more reddened sources, as indicated by the smaller [O~III]/[O~IV]
ratios show relatively
weaker O~VII. 

We determined the additional column density towards
the X-ray emission region in the Seyfert 2s and intermediate Seyferts
by, first, running a series of models with XSPEC 12 (Arnaud 1996), using tbabs, the Tuebingen-Boulder 
ISM absorption for the absorption model (Wilms, Allen and McCray 2000),
an O~VII line width $\sigma =3$ eV and assuming the line is observed at the 
emitted energy, in order to obtain the ratios of observed line
strength to the intrinsic line strength as a function of N$_{H}$. Then, we
determined N$_{H}$ for the individual sources by matching
the ratio of the observed 
O~VII-f/[O~IV] to the mean
value for the Seyfert 1s, O~VII-f/[O~IV] $=$ 0.31, to the computed
observed/emitted ratio.  
The values are listed in Table 5. In 7/10 sources, the X-ray column is either on the same order 
or somewhat smaller than the column of dusty gas. The latter can occur if the
absorber is ionized, hence the derived X-ray column would be a lower limit. For Mrk 3 and MCG$-$03$-$34$-$64, the X-ray
column is significantly larger than the dusty column. As noted in Section 3., Mrk 3 
may have a lower ISM dust-to-gas than in the Galactic ISM. MCG$-$03$-$34$-$64
shows evidence of Wolf-Rayet stars (Cid Fernandes et al. 2004), indicative of a young
stellar population, which may strengthen [O~III] relative to [O~IV], hence 
making the [O~III]/[O~IV] ratio a less reliable indicator of dust towards
the NLR. NGC 424 has [O~III]/[O~IV] greater than the Seyfert 1 mean, in keeping
with the possibility of a direct view of the inner NLR (e.g. Murayama et al. 1998),
however its O~VII-f/[O~IV] ratio indicates significant absorption. One possibility
is that much of the O~VII forms closer to the AGN than the [O~III] (see discussion in Iwasawa et al. 2001)
and is absorbed by gas that does not cover a significant portion of the inner NLR. Or, it
may simply be that the gas external to the inner NLR of NGC 424 has a low dust-to-gas ratio.   

Based on photoionization modeling of RGS spectra of NGC 3516 (Turner et al. 2003) and NGC 4151 (Armentrout et al. 2007),
there is evidence that the soft X-ray emission line gas has similar properties, in terms of ionization
state and column density, to those of warm absorbers (e.g., Crenshaw, Kraemer, \& George 2003).
Since the warm absorbers may originate as part of a disk wind (Kraemer et al. 2005; Krongold et al.
2007, 2010; Turner et al. 2008), which may expand and dissipate unless confined,
it is possible that the associated emission is centrally peaked, as suggested in the case
of NGC 424. If the lower R$_{\rm O3}$'s detected in Seyfert 2s are indicative of greater extinction towards
the inner NLR, one might expect a correlation between the R$_{\rm O3}$ and the O~VII-f/[O~IV] ratios.
However, for the eleven objects for which we have both [O~III] $\lambda$4363 and O~VII-f fluxes,
we obtain a Spearman rank coefficient of 0.49 and
significance of deviation from zero of 0.13, thus there is no strong evidence for a correlation.
This suggests that, unlike [O~III] $\lambda$4363, the soft X-ray emission is not typically centrally
peaked, although there are too few data points to make any definite conclusions.

\section{Distribution of Dusty Gas External to the NLR}

The lower [O~III]/[O~IV], O~VII-f/[O~IV] and R$_{\rm O3}$ ratios in Seyfert 2s, as compared to Seyfert 1s,
are consistent with a line-of-sight through dusty gas outside the optical NLR. The question
is how this gas is distributed and, specifically whether it is associated with the torus,
(e.g., Nagao et al. 2001; Baum et al. 2010) or dust structure at larger radial distances (e.g,
Capetti et al. 1996; Malkan, Gorjian, \& Tam 1998; also, see discussion below).
Krolik \& Begelman (1988) determined that the ratio of the scale height of the torus to its radius 
is $\sim $ 0.7. Based on observations of H$_{2}$O MASERs, the outer radius is $\sim$ 2 -- 15 pc 
(Gallimore et al. 1996; Taniguchi \& Murayama 1998), in general agreement with
IR interferometric observations of nuclear dust distributions in several
nearby Seyferts (Tristram et al. 2007; Raban et al. 2009; Burtscher et al.
2009). Therefore, if emission that arises
from larger radial distances is absorbed, it must be by dusty gas more distant
than the outer radii determined from the H$_{2}$O MASERs. 

The correlation between the O~VII-f/[O~IV] 
and [O~III]/[O~IV] ratios is generally
consistent with a scenario in which the O~VII and [O~III] regions are 1) co-located and 2) 
viewed through the same dusty gas in Seyfert 2 galaxies, but it does not lead to clear constraints
on the distribution of the external gas. 
Based on our modeling of NGC 4151 (Crenshaw \& Kraemer 2005), assuming a density law $n_{H} \propto r^{-1.7}$
(Kraemer et al. 2000), the radial distance at which $n_{H}$ falls to 10$^{4.8}$ cm$^{-3}$, or
an order of magnitude below the critical density for de-excitation of the $^{1}D_{2}$ level of O~III
(Osterbrock \& Ferland 2006), is $\sim$ 13 pc. Hence, it is plausible that the dense O~III-emitting
component could be hidden if our view of NGC 4151 intersected the torus, as in Seyfert 2s. The radial
profile of [O~III] $\lambda$5007, Figure 3, provides
additional constraints on the distribution of the dust (for the details on the analysis, see
Crenshaw \& Kraemer 2005). The emission
is centrally peaked, as seen in other Seyfert 1s (e.g., Schmitt et al. 2003a), but only $\sim$ 30\%
of the total [O~III] comes from within $\sim$ 30 pc of 
the central source. Therefore, even the
complete extinction of this component would be insufficient to account for the difference
in the [O~III]/[O~IV] ratios of Seyfert 1s and 2s. If Seyfert 2 galaxies have similar
intrinsic [O~III] profiles, the dust must extend further out along the NLR. 

One way to test if the external dust is thickest towards the apex of the bicone is to compare
the extinction of the dense [O~III] component with the overall extinction of [O~III] $\lambda$5007.
Based on Nagao et al. (2001), the differences in the contribution of the dense component
in Seyfert 1s and 2s typically requires $\sim$ 80\% suppression of [O~III] $\lambda$4363 in the latter.
For an ISM extinction curve and dust-to-gas ratio, this corresponds to
an N$_{H}$ $\approx$ 2.2 $\times$ 10$^{21}$ cm$^{-2}$, which is within the 95\%
confidence interval for N$_{H}$ computed from the [O~III]/[O~IV] ratios (see
Section 3). If the [O~III] $\lambda$5007 is extended, as the STIS spectrum of NGC 4151 indicates, this
suggests that the dust is not centrally peaked, but,
rather, extends further along the NLR. Also, based on photoionization modeling of STIS longslit
spectra of the Seyfert 2 galaxies NGC 1068 (Kraemer \& Crenshaw 2000b),
Mrk 3 (Collins et al. 2009), and Mrk 573 (Kraemer et al. 2009), we have found that the
innermost optically-detected emission to be 10's of parsecs from the AGN, which is
consistent with extended dust structure.

Malkan et al. (1998) found evidence for excess galactic dust in
Seyfert 2 galaxies, at
radial distance of up to several 100s of pc, which could obscure the BLR and
central continuum. They attributed the absence of such structures in
Seyfert 1s as due to differences in the host galaxy properties. However,
de Zotti \& Gaskell (1985) suggested that gas associated with the host
galaxy is evacuated out to distances of $\lesssim$ 100 pc, presumably due to
the effect of radiation pressure and/or winds from the AGN. As discussed in
Crenshaw \& Kraemer (2001), such a scenario would lead to differences in
reddening as a function of viewing angle towards the AGN. When one's
line-of-sight is close to the bicone axis, i.e. when viewing the object as a
Seyfert 1, there is little extinction. However, when the bicone axis is closer to
the plane of the sky, i.e., when the object is seen as Seyfert 2, the intervening gas
has been less affected by the AGN, hence the extinction
can be significant. The fact that there is evidence for dust external to the optical NLR in NGC 1068 
(Kraemer \& Crenshaw 2000a), but not in the case of NGC 4151 (Kraemer et al. 2000),
is consistent with this scenario. Furthermore, Capetti et al. (1996) estimate
a scale height of $\sim$ 180 pc for the dust lane crossing the nucleus in the Seyfert 2 
galaxy Mrk 78. In summary, there is evidence that obscuring dust extends
far along the NLR. Models of extended (e.g., Garanato, Danese, \& Franceschini 1997) or clumpy
tori (Nenkova, Ivezi\'c, \& Elitzur 2002) suggest that associated dust structure may 
extend $\sim$ 100 pc. Another possibility is that the dust structure is associated with
nuclear dust spirals (Regan \& Mulchaey 1999; Pogge \& Martini 2002; Martini
et al. 2003b), which, in turn may be part of the fueling flow to the AGN
(Martini al. 2003a; Deo, Crenshaw, \& Kraemer 2006; Sim\~{o}es Lopes et al. 2007).

\section{Summary} 

We have selected 108 AGN from the RSA, 12$\mu$m, and BAT samples,
after rejecting those in peculiar host galaxies, to study the
effects of extinction and absorption towards the NLR. 
First, we compared ground-based [O~III] $\lambda$5007 fluxes, taken
from the literature, with [O~IV] 25.89$\mu$m fluxes from
{\it Spitzer}/IRS observations.  Using [O~IV]
as a proxy for bolometric luminosity, we find evidence for
larger [O~III]/[O~IV] in the least luminous sources, which
we suggest is an ionization affect (e.g., Netzer 2009). 
We also find that AGN in host galaxies with $b/a < 0.5$ tend
to have lower [O~III]/[O~IV] ratios, which we attribute to
extinction by dust in the plane of the host galaxy. After eliminating
the inclined objects from our sample, we were left with
40 Seyfert 2s and 26 Seyfert 1s. We find the following:

1. In the reduced sample, the [O~III]/[O~IV] ratio is lower, with statistical significance, 
in Seyfert 2s than Seyfert 1s, which is consistent
with the NLR is Seyfert 2s being viewed though an additional
column of dusty gas with N$_{H}$ $\sim$ 2 $\times$ 10$^{21}$ cm$^{-2}$. Although the
relative weakness of [O~III] in Seyfert 2s has been noted
previously (e.g., Mel\'endez et al. 2008a; Diamond-Stanic et al. 2009;
Baum et al. 2010, Lamassa et al. 2010), this is first time the effect
has been demonstrated after accounting for possible effects from
the host galaxy. We find that intermediate Seyferts have
[O~III]/[O~IV] ratios spanning the range seen in Seyfert 1s and 2s,
which suggests that some may have weak BLR emission due to reddening
while others may possess weak AGN, hence intrinsically weak BLR emission (see Trippe et al. 2010). 

2. We compared the [O~III]/[O~IV] ratios to R$_{\rm O3}$, for a subset of the sample for which [O~III] $\lambda$4363 measurements were
available: 20 Seyfert 2s, 21 Seyfert 1s, and 2 intermediate Seyferts. Seyfert 2s were generally found to have both low [O~III]/[O~IV] ratios and R$_{\rm O3}$
below the low-density limit for gas at 20,000K. As shown by Nagao et al. (2001), most Seyfert 1s have
R$_{\rm O3}$ above the low-density limit, consistent with a contribution from gas with $n_{H} > 10^{6}$ cm$^{-3}$,
which is likely within $\sim$ 15 pc of the central source (e.g. Crenshaw \& Kraemer 2005).
If extinction reduces the contribution to [O~III] $\lambda$ 4363 from the dense
component by 80\% in the Seyfert 2s, as suggested by Nagao et al., this region
would be viewed though dusty gas with column density similar to the mean value derived
from the [O~III]/[O~IV] ratios.  

3. Limited by the availability of high-resolution X-ray spectra, we were able to compare the ratios 
of O~VII-f/[O~IV] and [O~III]/[O~IV] for 17 Seyfert in our sample. We find a correlation
between the ratios, which is expected given the evidence that O~VII-f and [O~III] regions are
co-located (e.g. Zeng 2009). We estimated the amount of intervening material required
to account for the absorption of the O~VII emission in Seyfert 2s and find, with a few
exceptions, that
the X-ray absorbing column density are roughly on the same order as those of the
dusty gas. While lower X-ray predicted column densities, compared to dusty columns, may
indicate that
the absorbers are partially ionized, the higher X-ray predicted columns are more consistent with low
dust-to-gas ratios in the absorbers. 

4. Using a STIS G430M slitless spectrum of NGC 4151, we determined that only $\sim$ 30\% of the 
[O~III] $\lambda$ 5007 comes from the inner $\sim$ 30 pc. If Seyfert 2s have similar
intrinsic [O~III] profiles, the dust must extend to larger radial distances. 
Based on this and the results of our previous photoionization modeling of Seyfert
2 galaxies, we suggest that dusty gas surrounds the NLR and extends out to radial distances
of at least 10s of parsecs. Based on scale-height constraints for torus models, the obscuring
gas must lie further from the AGN than the maximum radial distance estimated from H$_{2}$O MASERs.  
One possibility is this gas is associated with the nuclear dust spirals observed in
many Seyferts. In any case, it points to the presence of a significant amount of material
outside the optical NLR, in agreement with results from NIFS spectra of a several
Seyfert galaxies (Riffel et al. 2008; Storchi-Bergmann et al. 2009;
Riffel, Storchi-Bergmann, \& Nagar 2010).

 
\acknowledgments
Basic research in Astronomy at the NRL is supported by 6.1 base funding.
This research has made use of the NASA/IPAC Extragalactic Database (NED)
which is operated by the Jet Propulsion Laboratory, California Institute of 
Technology, under contract with the National Aeronautics and Space 
Administration. We thank Juliette Buet for her assistance with this project.
We thank an anonymous referee for valuable suggestions.

\clearpage

\figcaption[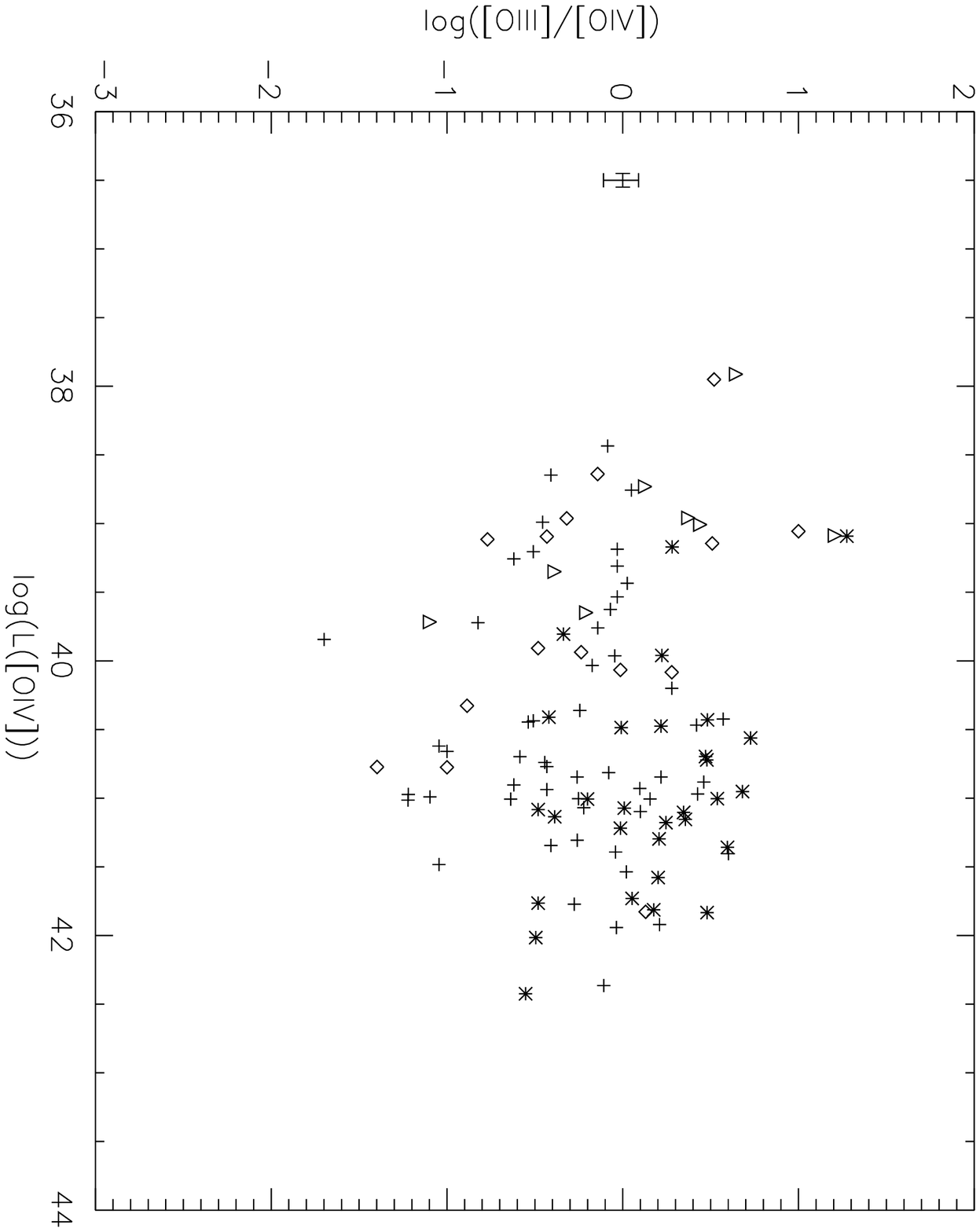]{The ratio of [O~III]$\lambda$5007/[O~IV] 25.89$\mu$m 
plotted against [O~IV] luminosity, for the complete sample. Symbols are
as follows: Seyfert 1s, asterisks; Seyfert 2s, crosses; Intermediate Seyferts,
diamonds; LINERs, triangles. Typical uncertainty is shown at the left of the
figure. As noted in Section 2, while there is no overall trend, a number of lower luminosity
objects have large [O~III]/[O~IV] ratios, which we attribute to an
ionization effect.}

\figcaption[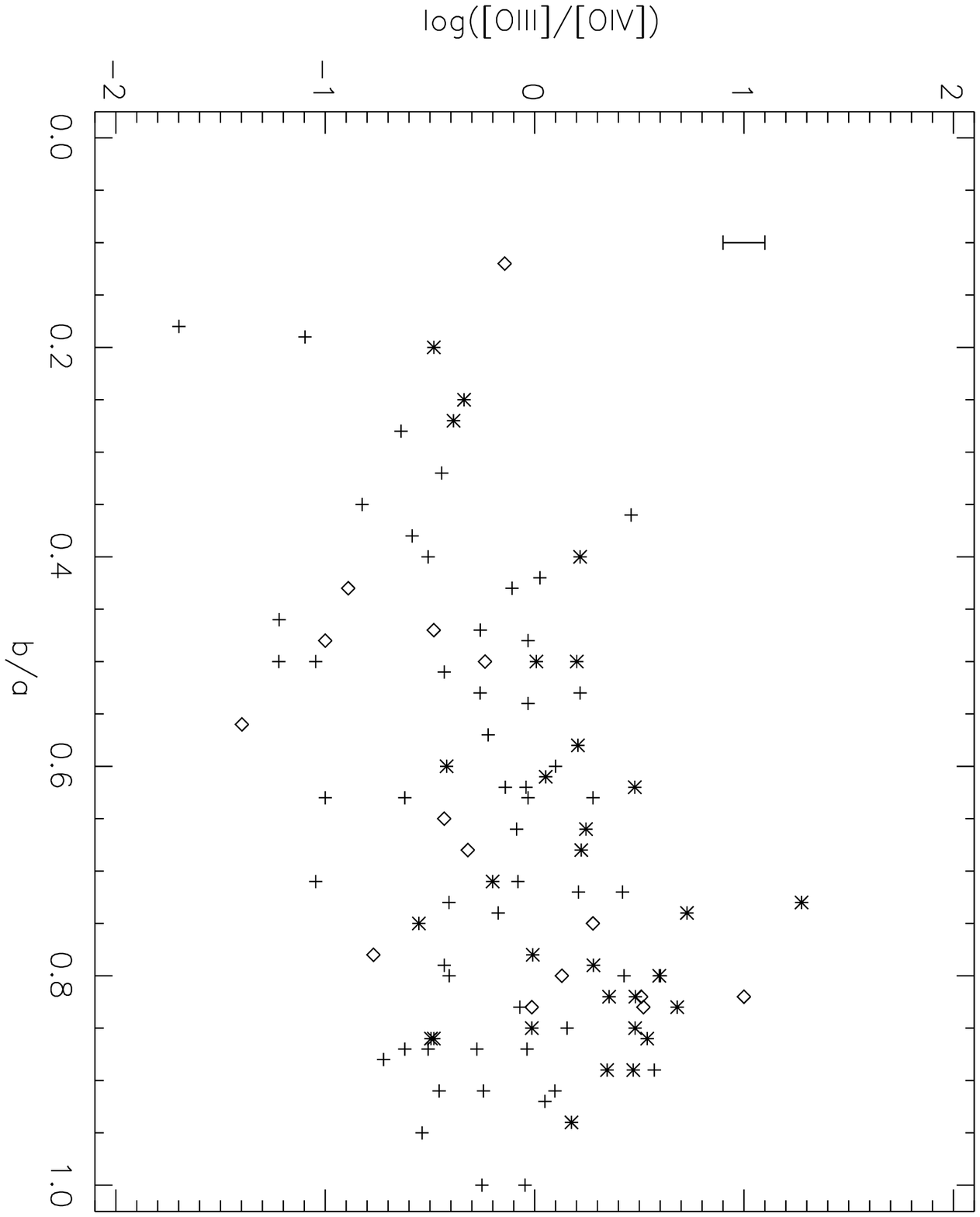]{The [O~III]/[O~IV] ratio as a function of inclination (the
ratio of the semi-minor to semi-major axis of the host galaxy). Symbols are as
in Figure 1, and the typical uncertainty in [O~III]/[O~IV] is shown at the left.
Note the relative paucity of sources with [O~III]/[O~IV] greater than unity in hosts with
$b/a < 0.5$.}

\figcaption[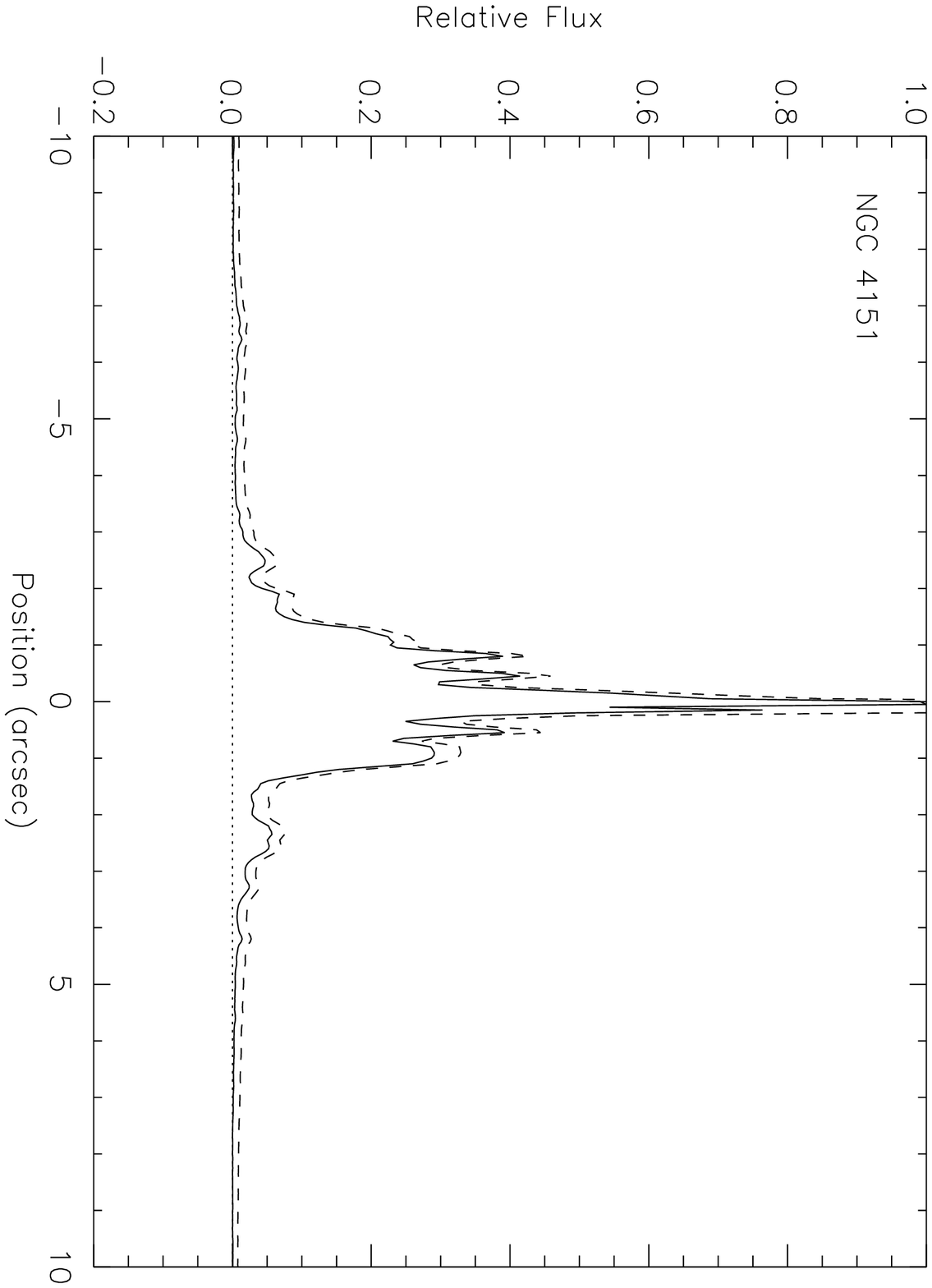]{Plot of the surface brightness, co-added in the dispersion 
direction, as a function of distance from the nucleus in the cross-dispersion 
direction in a STIS/G430M slitless spectrum of NGC 4151. The top profile is 
continuum plus [O~III] emission, and the bottom profile is just the [O III] 
emission, obtained by subtracting the continuum profile from a nearby region.}

\figcaption[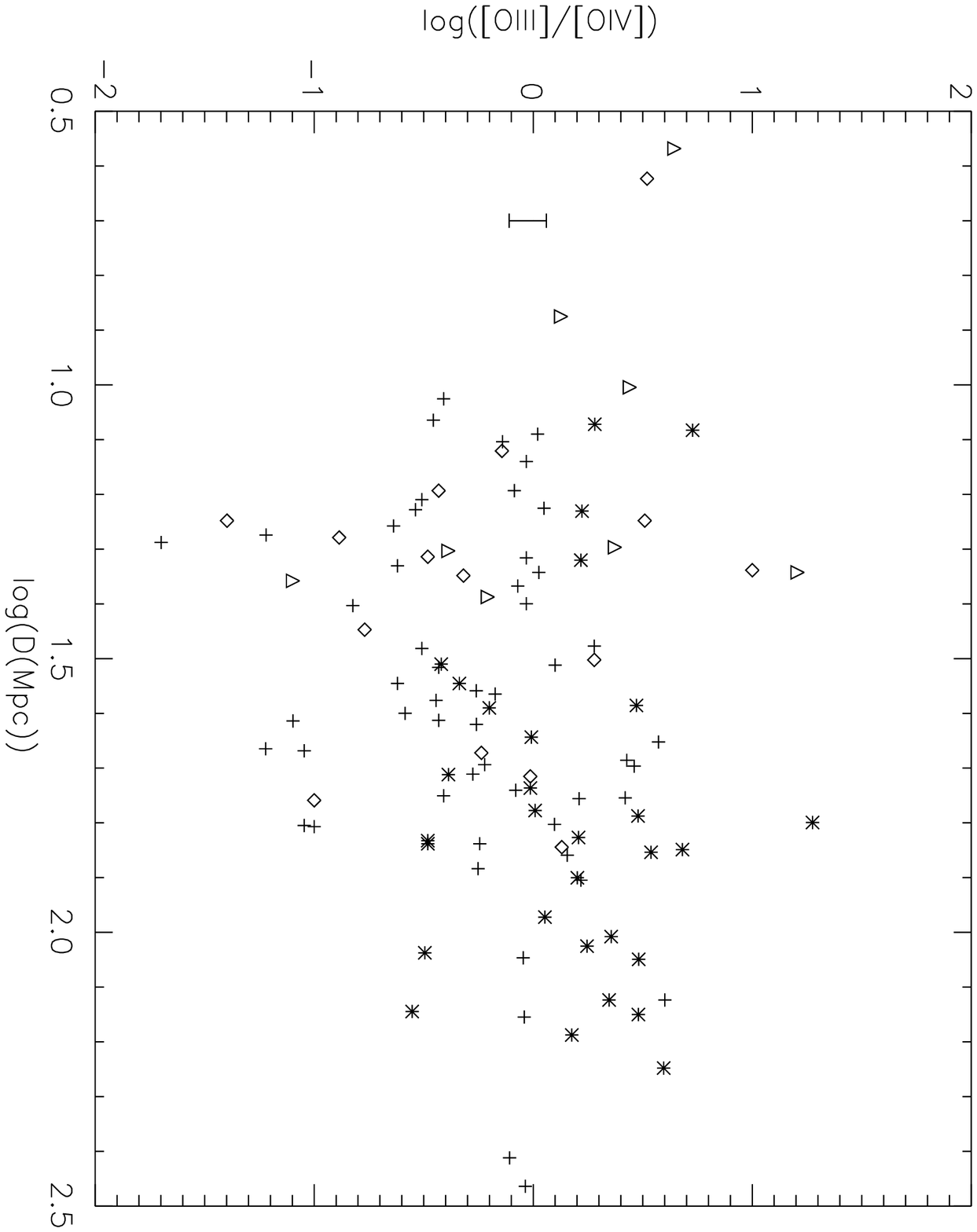]{The [O~III]/[O~IV] ratio as a function of distance
(distances are mean values from the literature, as listed in NED; if none are
available, distances are determined from $z$, assuming H$_{\rm o} = 71$ km s$^{-1}$
Mpc$^{-1}$). Symbols as in Figure 1, with the typical uncertainty in 
[O~III]/[O~IV] shown at left.}

\figcaption[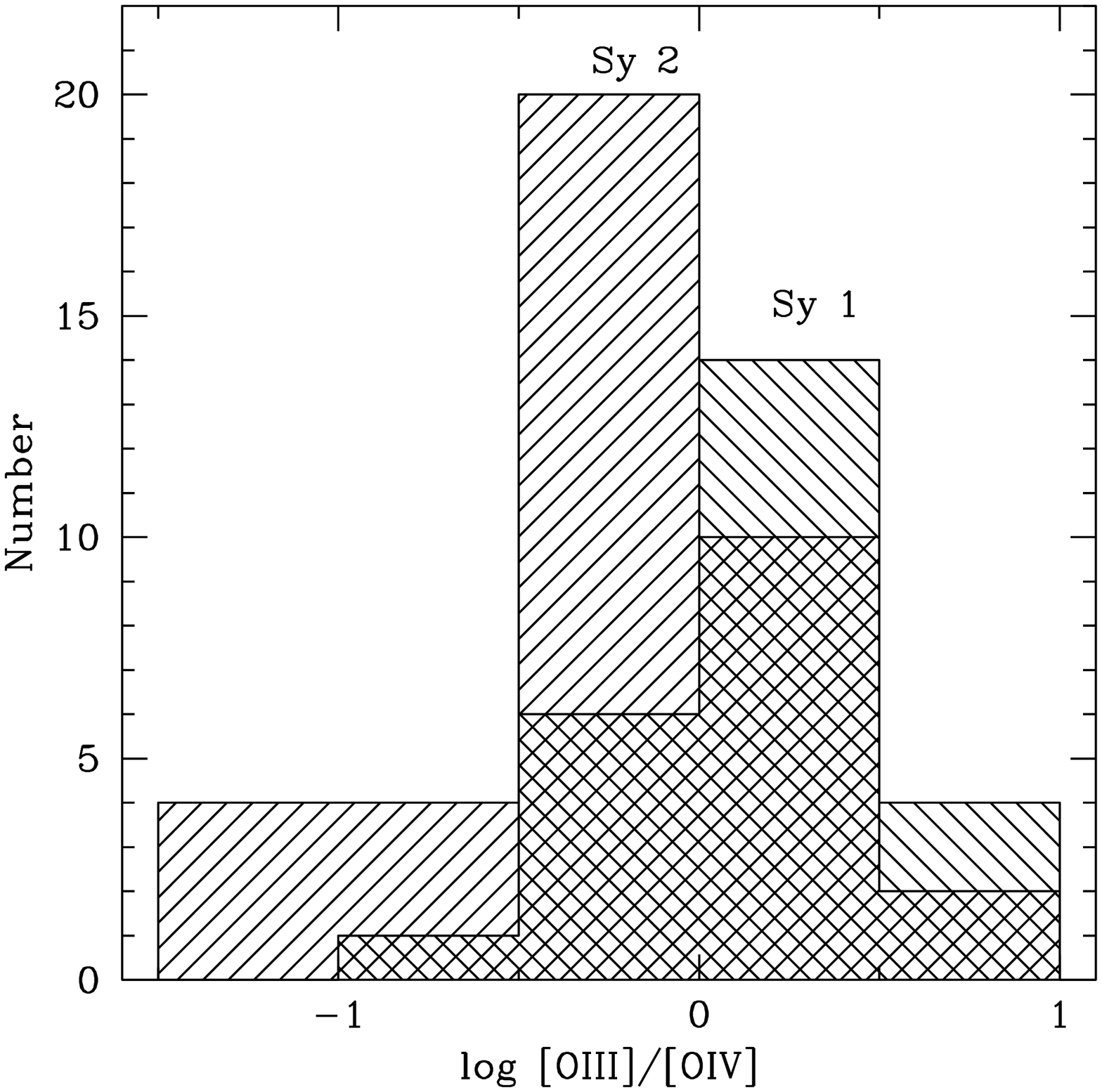]{Histogram showing the [O~III]/[O~IV] ratios for the
Seyfert 1s and Seyfert 2s in the reduced sample (see discussion in Sections 2
and 3).}

\figcaption[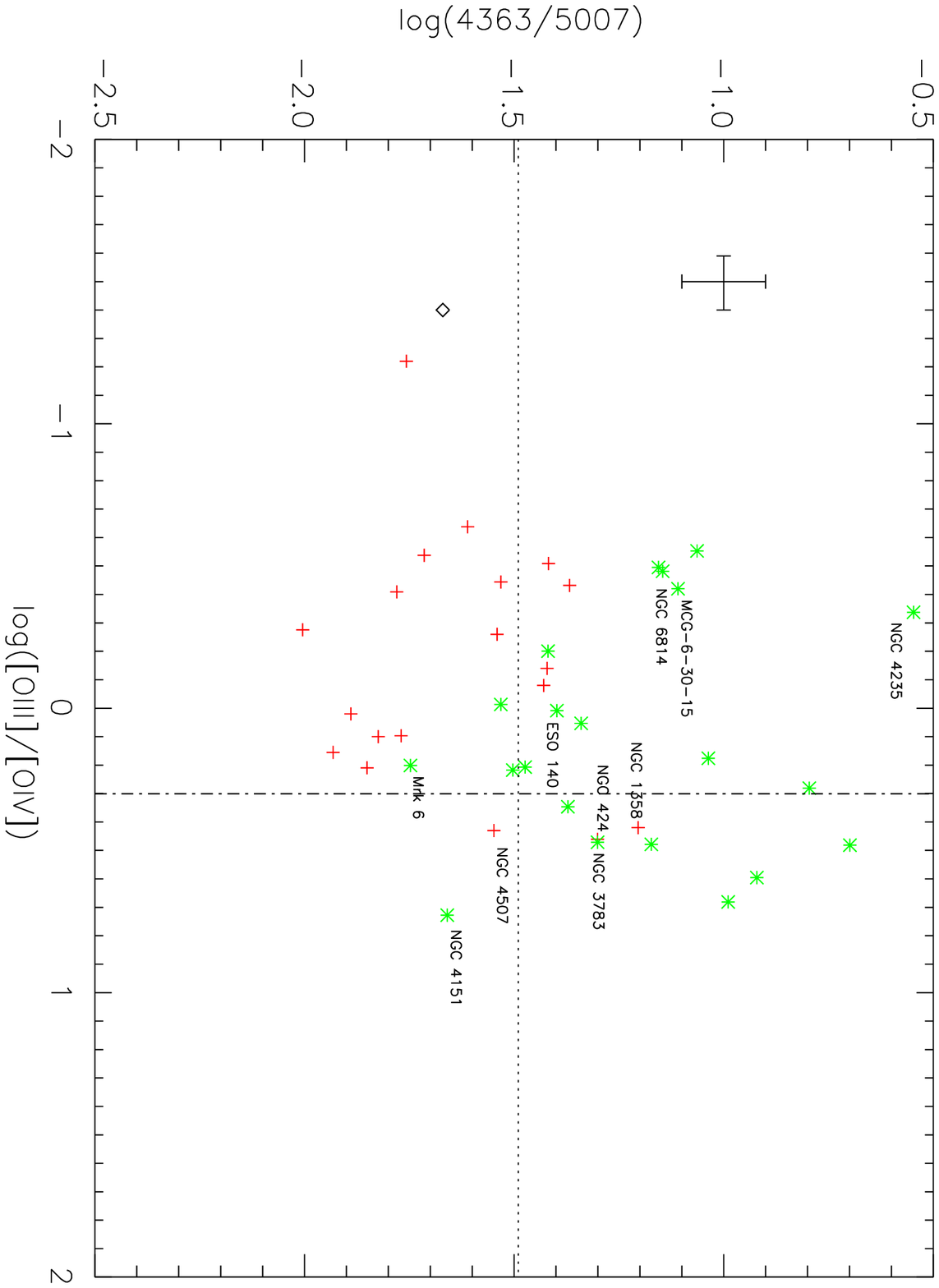]{Comparison of R$_{\rm O3}$ and [O~III]/[O~IV]. Symbols as
in Figure 1, except Seyfert 1s are shown in green and Seyfert 2s are shown in red; typical uncertainty in the ratios are
shown at left. Several objects discussed in Section 4.1 are identified. The
vertical dash-dotted line shows the mean [O~III]/[O~IV] ratio for the Seyfert 1s
in the reduced sample, while the horizontal dotted line shows the value of 
R$_{\rm O3}$ for a 20,000K in the low-density limit.}

\figcaption[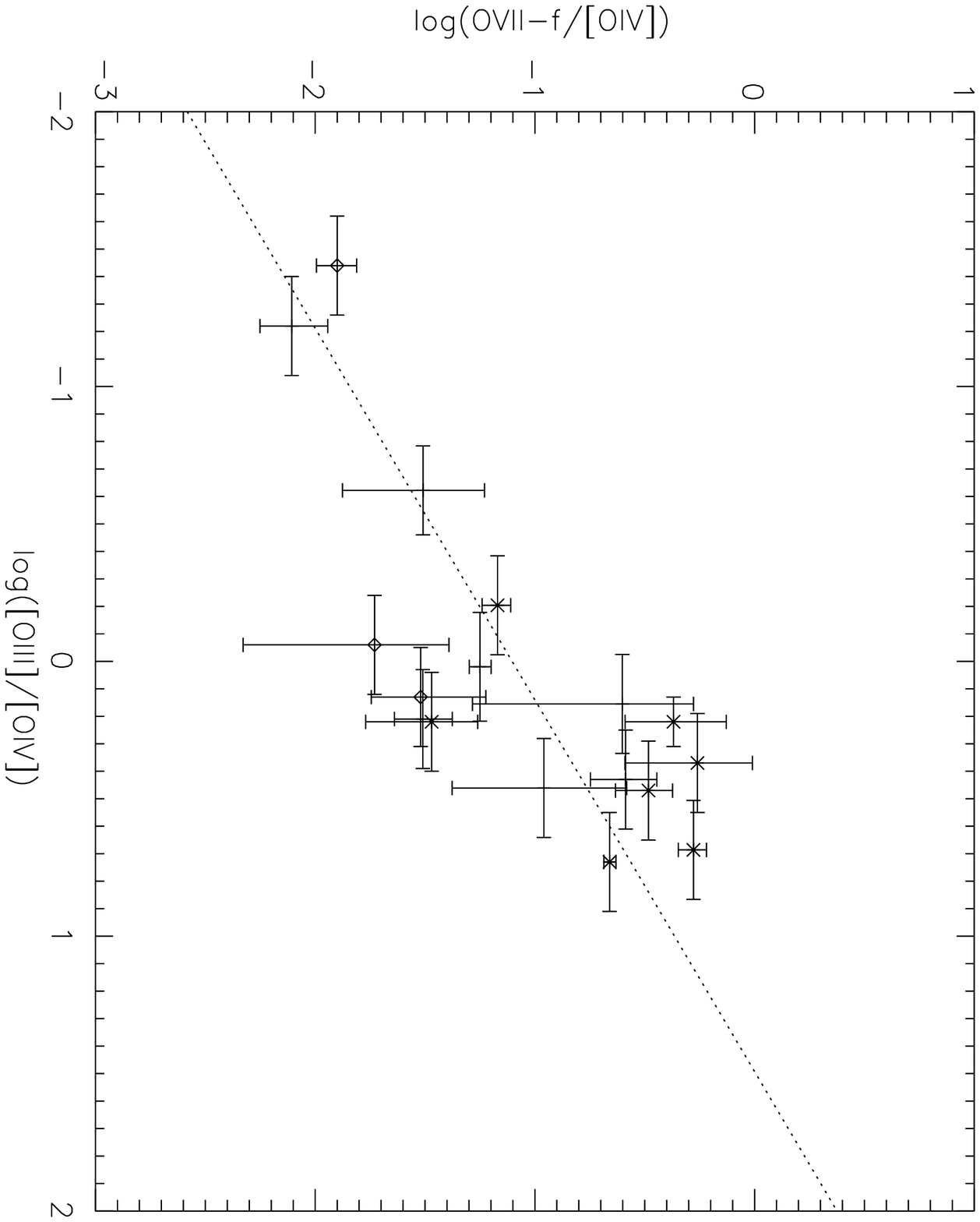]{Comparison of the OVII-f/[O~IV] and [O~III]/[O~IV] ratios.
Symbols as in Figure 1. The over-plotted line shows a linear
regression fit to the points.}

\newpage

\begin{deluxetable}{lllllll}
\tabletypesize{\scriptsize}
\tablecaption{Properties of the Seyfert 2 Galaxies in the Sample$^{a}$ \label{tbl-1}}
\tablewidth{0pt}
\tablehead{
\colhead{name} & \colhead{Distance(Mpc)$^{b}$} & \colhead{original sample(s)} & \colhead{F[O IV]$^{c}$}
&\colhead{F[OIII]$^{c}$} &\colhead{b$/$a} &\colhead{logR$_{\rm O3}$}
}
\startdata
Mrk 348 & 63.5 & $12\mu$m, BAT & 1.76$\times10^{-13}$; 1 & 2.2$\times 10^{-13}$; 2 & 0.91, 3 & $-1.77$; 4\\
IRAS 00521$-$7054 & 291.0 & $12\mu$m & 8.63 $\times 10^{-14}$; 1 & 7.9 $\times 10^{-14}$; 2 & 0.87; 5 & \\
NGC 424 & 49.7 & $12\mu$m & 2.58 $\times 10^{-13}$; 6 & 7.5 $\times 10^{-13}$; 2 & 0.36; 3 & $-1.30$; 4 \\
NGC 513 & 80.3 & $12\mu$m, RSA, BAT  & 9.09 $\times 10^{-14}$; 6 & 1.5 $\times 10^{-13}$; 2 & 0.53; 3 & \\
IRAS01475$-$0740 & 23.3 &  $12\mu$m & 6.49 $\times 10^{-14}$; 1 & 5.5 $\times 10^{-14}$; 2 & 0.83; 5 & \\
NGC 788 & 55.0 &  RSA, BAT & 1.80 $\times 10^{-13}$; 6 & 1.5 $\times 10^{-13}$; 17 & 0.71; 3 & $-1.43$; 4 \\
NGC 1068 & 12.3 & RSA & 1.90$\times 10^{-11}$; 8$^{d}$ & 2.0 $\times 10^{-11}$; 2 & 0.88; 9 & $-1.89$; 4\\
NGC 1125 & 46.2 & $12\mu$m & 4.04$ \times 10^{-13}$; 6 & 2.3 $\times 10^{-14}$; 7 & 0.50; 5 & \\
ESO 417$-$G006 & 68.8 & BAT & 4.04$ \times 10^{-14}$; 10 & 2.3 $\times 10^{-14}$; 11 & 0.91; 5& \\
Mrk 1066 & 49.4 & RSA & 4.02 $\times 10^{-13}$; 10 & 2.4 $\times 10^{-13}$; 2 & 0.57; 9 & $-1.65$; 12 \\
NGC 1320 & 37.7 & $12\mu$m, RSA & 3.23 $\times 10^{-13}$; 1 & 1.2 $\times 10^{-13}$; 2 & 0.32; 5 & $-1.53$; 4 \\
NGC 1358 & 56.8 & RSA & 7.61 $\times 10^{-14}$; 8 & 2.0 $\times 10^{-13}$; 8 & 0.72; 3 & $-1.20$; 4 \\
NGC 1386 & 16.2 & $12\mu$m, RSA & 8.70 $\times 10^{-13}$; 8 & 2.7 $\times 10^{-13}$; 8 & 0.40; 3 & $-1.42$; 4 \\
NGC 1433 & 11.6 & RSA & 6.07 $\times 10^{-14}$; 8 & 2.1 $\times 10^{-14}$; 8 & 0.91; 5 & \\
F04385$-$0828 & 63.8 &  $12\mu$m & 8.56 $\times 10^{-14}$; 1 & 8.0 $\times 10^{-15}$; 2  & 0.50; 9 & \\
NGC 1667 & 64.1 &  $12\mu$m, RSA & 9.28 $\times 10^{-14}$; 8  & 9.1 $\times 10^{-15}$; 8 & 0.63; 3 & \\
ESO 033$-$G02 & 76.5 &  $12\mu$m, BAT  & 1.44 $\times 10^{-13}$; 6 & 8.1 $\times 10^{-14}$; 2 & 1.0; 5 & \\
NGC 2110 & 32.8 & RSA, BAT & 4.57 $\times 10^{-13}$; 10 & 1.7 $\times 10^{-13}$; 2 & 0.79; 3  & $-1.37$; 4 \\
Mrk 3 & 57.0 &  BAT & 2.14 $\times 10^{-12}$; 10 & 3.5 $\times 10^{-12}$; 13 & 0.72; 3 & $-1.85$; 4 \\
NGC 2273 & 30.0 &  RSA & 1.47 $\times 10^{-14}$; 8 & 2.8 $\times 10^{-13}$; 8 & 0.63; 3 & \\
MGC$+$01$-$24$-$012 & 142.8 & BAT & 1.01 $\times 10^{-13}$; 10 & 9.2 $\times 10^{-14}$; 2 & 0.62; 3 &\\
NGC 3081 & 32.5 &  RSA, BAT & 9.89 $\times 10^{-13}$; 8 & 1.3 $\times 10^{-12}$; 17 & 0.60; 3 & $-1.82$; 4 \\
NGC 3079 & 19.4 & RSA, BAT & 1.53 $\times 10^{-13}$; 8 & 1.6 $\times 10^{-15}$; 8 & 0.18; 5& \\
IC 2560 & 35.1 & RSA & 5.43 $\times 10^{-13}$; 8  & 1.3 $\times 10^{-13}$; 8 & 0.63; 5 & \\
NGC 3185 & 22.0 & RSA & 4.70 $\times 10^{-14}$; 8 & 5.0 $\times 10^{-14}$; 8 & 0.42; 5 & \\
NGC 3393 & 51.4 & RSA & 1.87 $\times 10^{-12}$; 10 & 9.9 $\times 10^{-13}$; 2 & 0.87; 9  & $-2.00$; 4 \\
NGC 3486 & 10.6 & RSA & 3.30 $\times 10^{-14}$; 8 & 1.3 $\times 10^{-14}$; 8 & 0.72; 3 & \\
NGC 3735 & 41.1 & RSA & 4.84 $\times 10^{-13}$; 8 & 3.7 $\times 10^{-14}$; 8 & 0.19; 5 & \\
NGC 3941 & 15.6 & RSA & 9.35 $\times 10^{-15}$; 8 & 7.7 $\times 10^{-15}$; 8 & 0.66; 5 & \\
NGC 4388 & 18.1 & $12\mu$m, RSA, BAT & 2.59 $\times 10^{-12}$; 8  & 5.9 $\times 10^{-13}$; 2  & 0.28; 9 & $-1.61$; 4\\
NGC 4477 & 16.8 & RSA & 1.69 $\times 10^{-14}$; 8 & 1.9 $\times 10^{-14}$; 8 & 0.92; 5 & \\
NGC 4501 & 20.7 & $12\mu$m, RSA & 3.98 $\times 10^{-14}$; 8 & 3.7 $\times 10^{-14}$; 8 & 0.54; 5 &\\
NGC 4507 & 48.5 & RSA, BAT & 3.31 $\times 10^{-13}$; 8 & 8.9 $\times 10^{-13}$; 2 & 0.80; 3 & $-1.55$; 4 \\
Tol 1238$-$364 & 44.9 & $12\mu$m, RSA & 1.1 $\times 10^{-13}$; 6 & 4.1 $\times 10^{-13}$; 14 & 0.89; 3 & \\
NGC 4698 & 41.7 & RSA & 2.04 $\times 10^{-14}$; 8 &  1.9 $\times 10^{-14}$; 8 & 0.63; 5 & \\
NGC 4941 & 13.8 & $12\mu$m, RSA & 1.50 $\times 10^{-13}$; 8 & 1.4 $\times 10^{-13}$; 8 & 0.48; 3 & \\
NGC 4939 & 41.0 & RSA & 4.30 $\times 10^{-13}$; 8 & 1.6 $\times 10^{-13}$; 8 & 0.51; 5 & \\
NGC 4968 & 25.1 & $12\mu$m & 3.37 $\times 10^{-13}$; 6 & 1.9 $\times 10^{-13}$; 15 & 0.47; 5 & \\
NGC 5135 & 56.3 & $12\mu$m, RSA & 5.83 $\times 10^{-13}$; 8 & 2.3 $\times 10^{-13}$; 2 & 0.80; 3 & $-1.78$; 4 \\
NGC 5347 & 36.7 & $12\mu$m, RSA & 6.7 $\times 10^{-14}$; 1 & 4.5 $\times 10^{-14}$; 2 & 0.74; 3& \\
NGC 5631 & 30.3 & RSA & 1.46 $\times 10^{-14}$; 8  & 4.5 $\times 10^{-15}$; 8 & 0.87; 5 & \\
NGC 5643 & 16.9 & RSA & 8.16 $\times 10^{-13}$; 8  & 2.4 $\times 10^{-13}$; 8 & 0.95; 3 & $-1.71$; 4 \\
NGC 5728 & 36.2 & RSA, BAT & 1.29 $\times 10^{-12}$; 8 & 7.1 $\times 10^{-13}$; 2 & 0.53; 3 & $-1.54$; 4 \\
NGC 5899 & 39.8 & RSA & 2.63 $\times 10^{-13}$; 8  & 6.9 $\times 10^{-14}$; 8 & 0.38; 5 & \\
NGC 6300 & 12.7 & RSA & 2.98 $\times 10^{-13}$; 8 & 2.2 $\times 10^{-13}$; 2 & 0.62; 3 & $-1.42$; 4 \\
IC 5063 & 46.6 & $12\mu$m, RSA, BAT & 1.17 $\times 10^{-12}$; 1 & 1.0 $\times 10^{-13}$; 2 & 0.71; 3 & $-1.70$; 16 \\
Mrk 897 & 111.3 & $12\mu$m & 6.20 $\times 10^{-15}$; 6 & 5.9 $\times 10^{-15}$; 7 & 1.0; 5 & \\
IRASF22017$+$0319 & 258.2 & $12\mu$m & 2.90 $\times 10^{-13}$; 6 & 2.3 $\times 10^{-13}$; 2 & 0.43; 5 & \\
MGC$-$03$-$58$-$7 & 132.9 & 12$\mu$m, RSA & 1.20 $\times 10^{-13}$; 1 & 4.8 $\times 10^{-13}$; 2  & 0.80; 5 & \\
NGC 7582 & 18.8 &  $12\mu$m, RSA, BAT  & 2.22 $\times 10^{-12}$; 8 & 1.3 $\times 10^{-13}$; 2 & 0.46; 3 & $-1.76$; 4 \\
NGC 7590 & 25.3 & $12\mu$m, RSA & 6.88 $\times 10^{-14}$; 8 & 1.1 $\times 10^{-14}$; 8 & 0.35; 5 & \\
NGC 7682 & 72.3 & BAT & 1.62 $\times 10^{-13}$; 10 & 2.3 $\times 10^{-13}$; 2 & 0.85; 3 & $-1.75$; 4\\
NGC 7743 & 21.4 & RSA & 3.30 $\times 10^{-14}$; 8 & 7.9 $\times 10^{-15}$; 8 & 0.87; 5 & \\  
\enddata
\tablenotetext{a}{Second number in column is the reference for the value, as follows: 1. Tommasin et al. (2010); 
2.Bonatto \& Pastoriza (1997); 3.  Kirhakos \& Steiner (1990); 4.  Nagao et al. (2001), and references therein; 5. NED; ;
6. Tommasin et al. (2008); 7.  Gu et al. (2006); 8. Diamond-Stanic et al. (2009); and references therein; 9. de Zotti \& Gaskell (1985);
10. Mel\'endez et al. (in prep.); 11. Stauffer (1982); 12.  Goodrich \&
Osterbrock (1983); 
13. Vaceli et al. (1997); 13. Ho et al. (1997); 14. Cid Fernandes et al. (2001); 16. Phillips, Charles \& Baldwin (1983);
Whittle (1992).}
\tablenotetext{b}{Distances are mean values from literature (see NED) or, if none are available, calculated from
redshift, assuming H$_{\rm o} = 71$ km s$^{-1}$ Mpc$^{-1}$.}
\tablenotetext{c}{Flux in units of ergs cm$^{-2}$ s$^{-1}$.}
\label{tbl2}
\end{deluxetable}

\newpage 

\begin{deluxetable}{lllllll}
\tabletypesize{\scriptsize}
\tablecaption{Properties of the Seyfert 1 Galaxies in the Sample \label{tbl-3}}
\tablewidth{0pt}
\tablehead{
\colhead{name} & \colhead{Distance(Mpc)$^{b}$} & \colhead{original sample(s)} & \colhead{F[O IV]$^{c}$}
&\colhead{F[OIII]$^{c}$} &\colhead{b$/$a} &\colhead{logR$_{\rm O3}$}
}
\startdata
Mrk 335 & 109.0 & $12\mu$m, BAT & 7.28 $\times 10^{-13}$; 1 & 2.3 $\times 10^{-13}$; 17 & 0.86; 3 & $-1.16$; 4 \\
Mrk 352 & 63.0 & BAT & 2.60 $\times 10^{-15}$; 10 & 4.9 $\times 10^{-14}$; 17 & 0.73; 3 & \\
F9 & 177.0 & BAT & 6.08 $\times 10^{-14}$; 10 & 2.4 $\times 10^{-13}$; 17 & 0.80; 3 & $-0.92$; 4 \\
Mrk 590 & 112.0 &  BAT & 1.79 $\times 10^{-14}$; 10 & 5.4 $\times 10^{-14}$; 2 & 0.82; 3 & $-0.70$; 4 \\
ESO 545$-$G-13 & 101.8  & $12\mu$m & 1.15 $\times 10^{-13}$; 6 & 2.6 $\times 10^{-13}$; 2 & 0.82; 5 & \\
NGC 931 &51.5 &  $12\mu$m, BAT & 4.30 $\times 10^{-13}$; 6 & 1.8 $\times 10^{-13}$; 2 & 0.27; 3 & \\
IC 1816 & 71.4 & BAT & 1.65 $\times 10^{-13}$; 10 & 5.7 $\times 10^{-13}$; 2  & 0.86; 5 & \\
NGC 1566 & 11.8 & RSA & 8.88 $\times 10^{-14}$; 8 &  1.7 $\times 10^{-13}$; 8 & 0.79; 9 & $-0.80$; 4 \\
3C 120 & 139.5 & $12\mu$m, BAT & 1.14 $\times 10^{-12}$; 1 & 3.2 $\times 10^{-13}$; 2 & 0.75; 5 & $-1.06$; 4 \\
Mrk 6 & 79.5 & $12\mu$m, BAT & 5.00 $\times 10^{-13}$; 6 & 7.9 $\times 10^{-13}$; 2 & 0.50; 3 & $-1.75$; 4 \\
Mrk 79 & 93.8 & $12\mu$m, BAT & 5.10 $\times 10^{-13}$; 1 & 5.8 $\times 10^{-13}$; 2 & 0.61; 3 & $-1.34$; 4 \\
IC 486 & 106.0 & BAT & 1.12 $\times 10^{-13}$; 10 & 2.0 $\times 10^{-13}$; 2 & 0.66; 5 & \\
NGC 3227 & 20.9 & RSA, BAT & 5.71 $\times 10^{-13}$; 8 & 9.4 $\times 10^{-13}$; 8 & 0.40; 3 & $-1.50$; 4 \\
NGC 3516 & 38.9 & $12\mu$m, RSA, BAT & 5.60 $\times 10^{-13}$; 8 & 3.5 $\times 10^{-13}$; 8 &  0.71; 3 & $-1.42$; 4 \\
NGC 3783 & 38.5 & RSA, BAT & 2.80 $\times 10^{-13}$; 8 & 8.3 $\times 10^{-13}$; 8 & 0.89; 3 & $-1.30$; 4 \\
NGC 4051 & 17.0 & $12\mu$m, RSA, BAT & 2.64 $\times 10^{-13}$; 8 & 4.4 $\times 10^{-13}$; 8 & 0.68; 9 & \\
NGC 4151 & 12.1 & RSA, BAT & 2.08 $\times 10^{-12}$; 8 & 1.1 $\times 10^{-11}$; 8 & 0.74; 9 & $-1.66$; 4 \\
NGC 4235 & 35.1 & RSA, BAT & 4.33  $\times 10^{-14}$; 8 & 2.0 $\times 10^{-14}$; 8 & 0.25; 9 & $-0.55$; 4 \\
Mrk 766 & 54.5 & $12\mu$m, RSA, BAT & 4.65 $\times 10^{-13}$; 1 & 4.5 $\times 10^{-13}$; 2 & 0.85; 9 & $-1.53$; 4 \\
NGC 4593 & 44.0 &  $12\mu$m, RSA, BAT & 1.32 $\times 10^{-13}$; 8 & 1.3 $\times 10^{-13}$; 8 & 0.78; 9 & \\
MCG$-$6$-$30$-$15 & 32.7 & $12\mu$m, BAT & 2.00 $\times 10^{-13}$; 1 & 7.5 $\times 10^{-14}$; 2 & 0.60; 9 & $-1.11$; 4 \\
IC 4329A & 68.0 & $12\mu$m, BAT & 1.05 $\times 10^{-12}$; 1 & 3.5 $\times 10^{-13}$; 2  & 0.20; 3 & \\
NGC 5548 & 70.6 &  $12\mu$m, RSA, BAT & 1.50 $\times 10^{-13}$; 1 & 7.3 $\times 10^{-13}$; 2 & 0.83; 9 & $-0.99$; 4 \\
Mrk 817 & 132.9 & $12\mu$m, BAT & 6.00 $\times 10^{-14}$; 6 & 1.3 $\times 10^{-13}$; 2 & 0.89; 3 & $-1.37$; 4 \\
Mrk 841 & 154.0 & BAT & 2.29 $\times 10^{-13}$; 10 & 3.4 $\times 10^{-13}$; 2 & 0.94; 3 & $-1.04$; 4 \\
ESO 140$-$G043 & 47.8 &  BAT & 2.75 $\times 10^{-13}$; 10 & 2.8 $\times 10^{-13}$; 17 & 0.50; 3 & $-1.40$; 4\\
NGC 6814 & 68.9 & RSA & 2.13 $\times 10^{-13}$; 8 & 7.0 $\times 10^{-14}$; 8 & 0.86; 3 & $-1.15$; 4 \\
NGC 6860 & 61.3 & $12\mu$m, BAT & 1.17 $\times 10^{-13}$; 1 & 3.5 $\times 10^{-13}$; 2 & 0.62; 5 & $-1.17$; 4 \\
Mrk 509 & 141.3 & $12\mu$m, RSA, BAT & 2.85 $\times 10^{-13}$; 8 & 8.6 $\times 10^{-13}$; 8 & 0.85; 9 & \\
NGC 7469 & 67.1 & RSA, BAT & 3.67 $\times 10^{-13}$; 8 & 5.9 $\times 10^{-13}$; 8 & 0.58; 3 & $-1.47$; 4 \\
\enddata
\tablenotetext{a}{References: same as in Table 1.}
\tablenotetext{b}{Distances are mean values from literature (see NED) or, if none are available, calculated from
redshift, assuming H$_{\rm o} = 71$ km s$^{-1}$ Mpc$^{-1}$.}
\tablenotetext{c}{Flux in units of ergs cm$^{-2}$ s$^{-1}$.}
\tablenotetext{d}{Flux from Infared Space Observatory/Short Wave Spectrometer data (Lutz et al. 2000).}
\end{deluxetable}

\newpage 

\begin{deluxetable}{lllllll}
\tabletypesize{\scriptsize}
\tablecaption{Properties of the Seyfert 1.8 and 1.9  Galaxies in the Sample$^{a}$ \label{tbl-5}}
\tablewidth{0pt}
\tablehead{
\colhead{name} & \colhead{Distance(Mpc)$^{b}$} & \colhead{original sample(s)} & \colhead{F[O IV]$^{c}$}
&\colhead{F[OIII]$^{c}$} &\colhead{b$/$a} &\colhead{logR$_{\rm O3}$}
}
\startdata
NGC 1194 & 57.4 & $12\mu$m, BAT & 1.51 $\times 10^{-13}$; 1 & 1.5 $\times 10^{-14}$; 2 & 0.48; 5 & \\
NGC 1365 & 17.7 & $12\mu$m, RSA, BAT & 1.58 $\times 10^{-12}$; 8 & 6.2 $\times 10^{-14}$; 8 & 0.56; 3 & \\
NGC 2639 & 47.0 & $12\mu$m, RSA & 3.27 $\times 10^{-14}$; 8 & 1.9 $\times 10^{-14}$; 8 & 0.50; 3 & \\
NGC 2992 & 30.5  & RSA & 1.08 $\times 10^{-12}$; 8 & 9.5 $\times 10^{-13}$; 2 & 0.31; 5  & $-1.65$; 4 \\
NGC 3660 & 51.9 & $12\mu$m & 3.61 $\times 10^{-14}$; 6 & 3.5 $\times 10^{-14}$; 2 & 0.83; 3 & \\
NGC 3982 & 21.8 & $12\mu$m, RSA & 2.00 $\times 10^{-14}$; 1 & 2.0 $\times 10^{-13}$; 2 & 0.82; 3 & $-2.80$; 4 \\
NGC 4138 & 15.6 & RSA, BAT & 4.27 $\times 10^{-14}$; 8 & 1.6 $\times 10^{-14}$; 8 & 0.65; 5 & \\
NGC 4168 & 28.0 & RSA & 1.39 $\times 10^{-14}$; 8 & 2.4 $\times 10^{-15}$; 8 & 0.78; 5 & \\
NGC 4395 & 4.2 & RSA, BAT & 4.23 $\times 10^{-14}$; 8 & 1.4 $\times 10^{-13}$; 8 & 0.83; 5 & \\
NGC 4565 & 13.2 & RSA & 2.09 $\times 10^{-14}$; 8 & 1.5 $\times 10^{-14}$; 8 & 0.12; 5 & \\
NGC 4639 & 22.3 & RSA & 1.54 $\times 10^{-14}$; 8 & 7.5 $\times 10^{-15}$; 8 & 0.68; 5 & \\
NGC 5033 & 20.6 & RSA & 1.59 $\times 10^{-13}$; 8 & 5.3 $\times 10^{-14}$; 8 & 0.47; 3 & \\
MCG$-$03$-$34$-$64 & 70.0 &  $12\mu$m & 1.15 $\times 10^{-12}$; 1 & 1.6 $\times 10^{-12}$; 2 & 0.80; 5 & \\
NGC 5273 & 17.7 & RSA & 3.72 $\times 10^{-14}$; 8 & 1.2 $\times 10^{-13}$; 8 & 0.82; 9& \\
NGC 6890 & 31.8 & $12\mu$m, RSA & 1.00 $\times 10^{-13}$; 8 & 1.9 $\times 10^{-13}$; 8  & 0.75; 3 & \\
NGC 7314 & 19.0 & $12\mu$m, RSA, BAT & 4.91 $\times 10^{-13}$; 8 & 6.5 $\times 10^{-14}$; 2 & 0.43; 3 & $-2.67$; 4 \\
\enddata
\tablenotetext{a}{References: same as in Tables 1 and 2}
\tablenotetext{b}{Distances are mean values from literature (see NED) or, if none are available, calculated from
redshift, assuming H$_{\rm o} = 71$ km s$^{-1}$ Mpc$^{-1}$.}
\tablenotetext{c}{Flux in units of ergs cm$^{-2}$ s$^{-1}$.}
\end{deluxetable}

\newpage 

\begin{deluxetable}{llllll}
\tabletypesize{\scriptsize}
\tablecaption{Properties of the LINERS in the Sample$^{a}$ \label{tbl-6}}
\tablewidth{0pt}
\tablehead{
\colhead{name} & \colhead{Distance(Mpc)$^{b}$} & \colhead{original sample(s)} & \colhead{F[O IV]$^{c}$}
&\colhead{F[OIII]$^{c}$} &\colhead{b$/$a}
}
\startdata
NGC 2655 &  24.4 & RSA & 6.25$\times 10^{-14}$; 8 & 3.9$ \times 10^{-14}$; 8 & 0.84; 5 \\
NGC 3031 & 3.7 & RSA & 4.99$ \times 10^{-14}$; 8  & 2.2 $\times 10^{-13}$;  8 & 0.52; 5 \\
NGC 4258 & 7.9 &  RSA & 7.99 $\times 10^{-14}$; 8 & 1.0 $\times 10^{-13}$; 8 & 0.39; 5  \\
NGC 4579 & 16.8 & RSA & 2.83 $\times 10^{-14}$; 8  & 7.8 $\times 10^{-14}$; 8 & 0.80; 5  \\
NGC 5005 & 17.5 & $12\mu$m, RSA & 1.99 $\times 10^{-14}$; 8 & 4.7 $\times 10^{-14}$; 8 & 0.48; 5  \\
NGC 6951 & 14.1 & RSA & 8.37 $\times 10^{-14}$; 8 & 7.0 $\times 10^{-15}$; 8 & 0.82; 5  \\
NGC 7213 & 22.0 & $12\mu$m, RSA, BAT & 2.11 $\times 10^{-14}$; 8 & 3.4 $\times 10^{-13}$; 8 & 0.92; 3  \\
NGC 7410 & 20.1 & RSA & 4.63 $\times 10^{-14}$; 8 & 1.9 $\times 10^{-14}$; 8 & 0.31; 5  \\
\enddata
\tablenotetext{a}{References: same as in Tables 1 and 2.}
\tablenotetext{b}{Distances are mean values from literature (see NED) or, if none are available, calculated from
redshift, assuming H$_{\rm o} = 71$ km s$^{-1}$ Mpc$^{-1}$.}
\tablenotetext{c}{Flux in units of ergs cm$^{-2}$ s$^{-1}$.}
\end{deluxetable}

\newpage 

\begin{deluxetable}{lll}
\tabletypesize{\scriptsize}
\tablecaption{Estimated Hydrogen Column Densities \label{tbl-6}}
\tablewidth{0pt}
\tablehead{
\colhead{name} & \colhead{logN$_{H}$ (dust$^{a}$)} &\colhead{logN$_{H}$ (X-ray$^{a}$)}}
\startdata
Mrk 348	& 20.88 & \\
IRAS 00521$-$7054 & 21.11 & \\
NGC 424 & --$^{b}$ & 21.10 \\
NGC 513 & 20.49 & \\
IRAS01475$-$0740 & 21.15 & \\
NGC 788	& 21.15 & \\
NGC 1068 & 21.00 & 21.30 \\
NGC 1125 & 21.75 & \\
ESO417$-$G006	& 21.30 & \\
Mrk 1066 & 21.30 & \\
NGC 1365 & 21.81 & 21.60 \\
NGC 1433 & 21.45 & \\
F04385$-$0828 & 21.70 & \\
NGC 1667 & 21.69 & \\
ESO033$-$G02 & 21.32 & \\
NGC 2110 & 21.43 & \\
Mrk 3	& 20.53 & 21.45 \\
NGC 2273 & 19.92 & \\
MCG01$-$24$-$012 & 21.11 & \\
NGC 2992 & 21.11 & 21.54 \\
NGC 3081 & 20.88 & \\
IC 2560	 & 21.53 & 21.45 \\
NGC 3393 & 21.34 & \\
NGC 3486 & 21.43 & \\
NGC 3941 & 21.15 & \\
NGC 4477 & 20.97 & \\
NGC 4501 & 21.08 & \\
NGC 4507 &  & 20.11 \\
NGC 4698 & 21.08 & \\
NGC 4939 & 21.43 & \\
MCG$-$03$-$34$-$64 & 20.81 & 21.45\\
NGC 5135 & 21.43 & \\
NGC 5347 & 21.26 &\\
NGC 5631 & 21.48 & \\
NGC 5643 & 21.49 & \\
NGC 5728 & 21.32 & \\
NGC 6300 & 21.23 & \\
IC 5063	& 21.70 & \\
Mrk 897	& 21.11 & \\
NGC 7582 & 21.76 & 21.70\\
NGC 7682 & 20.74 & 20.20\\
NGC 7743 & 21.53 & \\
\enddata
\tablenotetext{a}{In units of cm$^{-2}$. For details, see text.}
\tablenotetext{b}{[O~III]/[O~IV] ratio $>$ mean for Seyfert 1s.}
\end{deluxetable}

\newpage 

\begin{deluxetable}{lll}
\tabletypesize{\scriptsize}
\tablecaption{O~VII $\lambda$ 22.1 \AA~ fluxes$^{a}$  \label{tbl-6}}
\tablewidth{0pt}
\tablehead{
\colhead{name} & \colhead{Flux} &\colhead{Reference}}
\startdata
ESO 362$-$G018 & 6.52$^{+5.23}_{-3.34}$ & 1 \\
IC 2560 & 1.70$^{+1.51}_{-0.97}$ & 1\\
MCG$-$03$-$34$-$64 & 3.48$^{+3.35}_{-1.37}$ & 1 \\
Mrk 3 & 6.62$^{+2.39}_{-1.68}$ & 1\\
NGC 1068 & 108.$^{+6.36}_{-6.11}$ & 1 \\
NGC 1365 & 2.16$^{+0.44}_{-0.41}$ & 1 \\
NGC 2992 & 2.01$^{+2.34}_{-1.50}$ & 1 \\
NGC 3227 & 1.96$^{+1.22}_{-0.96}$ & 1 \\
NGC 3516 & 3.78$^{+4.31}_{-3.25}$ & 2\\
NGC 3783 & 9.21$^{+2.49}_{-2.49}$ & 3 \\
NGC 4051 & 11.3$^{+3.32}_{-2.86}$ & 4 \\
NGC 4151 & 45.5$^{+2.30}_{-2.20}$ &  5 \\
NGC 424 & 2.83$^{+3.92}_{-1.74}$  & 1 \\
NGC 4507 & 8.57$^{+3.26}_{-2.58}$ & 1 \\
NGC 5548 & 7.91$^{+1.08}_{-1.08}$ & 6 \\
NGC 7582 & 1.36$^{+0.61}_{-0.52}$ & 1 \\
NGC 7682 & 4.05$^{+4.48}_{-3.20}$ & 1 \\
\enddata
\tablenotetext{a}{In units of 10$^{-14}$ ergs cm$^{-2}$ s$^{-1}$}
\tablenotetext{b}{1. This paper;  2. Turner et al. (2003); 3. Kaspi et al. (2002); 
Lobban et al. (in prep.); 5. Armentrout et al. (2007);
6. Steenbrugge et al. (2005).}
\end{deluxetable}

\clearpage
\begin{figure}
\plotone{fig1.ps}
\\Fig.~1
\end{figure}

\clearpage
\begin{figure}
\plotone{fig2.ps}
\\Fig.~2
\end{figure}

\clearpage
\begin{figure}
\plotone{fig3.ps}
\\Fig.~3
\end{figure}

\clearpage
\begin{figure}
\plotone{fig4.ps}
\\Fig.~4
\end{figure}

\clearpage
\begin{figure}
\plotone{fig5.ps}
\\Fig.~5
\end{figure}

\clearpage
\begin{figure}
\plotone{fig6.ps}
\\Fig.~6
\end{figure}

\clearpage
\begin{figure}
\plotone{fig7.ps}
\\Fig.~7
\end{figure}

\end{document}